\documentclass[sigconf]{acmart}
\AtBeginDocument{%
  \providecommand\BibTeX{{%
    \normalfont B\kern-0.5em{\scshape i\kern-0.25em b}\kern-0.8em\TeX}}}

\setcopyright{acmcopyright}
\copyrightyear{2018}
\acmYear{2018}
\acmDOI{XXXXXXX.XXXXXXX}
\acmConference[Conference acronym 'XX]{Make sure to enter the correct
  conference title from your rights confirmation emai}{June 03--05,
  2018}{Woodstock, NY}
\acmPrice{15.00}
\acmISBN{978-1-4503-XXXX-X/18/06}

\usepackage{subfigure}
\usepackage[ruled,linesnumbered]{algorithm2e}
\usepackage{enumitem}
\usepackage{multirow}
\usepackage{graphicx}
\usepackage{threeparttable}
\usepackage{booktabs}

\begin{document}

\title{Touch the Core: Exploring Task Dependence Among Hybrid Targets for Recommendation}

\author{Xing Tang}
\authornote{All authors contributed equally to this research.}
\affiliation{
  \institution{FiT, Tencent}
  \city{Shenzhen}
  \country{China}
}
\email{xing.tang@hotmail.com}

\author{Yang Qiao}
\authornotemark[1]
\affiliation{
  \institution{FiT, Tencent}
  \city{Shenzhen}
  \country{China}
}
\email{sunnyqiao@tencent.com}

\author{Fuyuan Lyu}
\authornotemark[1]
\authornote{This work is done when working at FiT Tencent.}
\affiliation{
  \institution{McGill University \& MILA}
  \city{Montreal}
  \country{Canada}
}
\email{fuyuan.lyu@mail.mcgill.ca}

\author{Dugang Liu}
\authornote{Corresponding Authors}
\affiliation{
  \institution{Guangdong Laboratory of Artificial Intelligence and Digital Economy (SZ)}
  \city{Shenzhen}
  \country{China}
}
\email{dugang.ldg@gmail.com}

\author{Xiuqiang He}
\authornotemark[3]
\affiliation{
  \institution{FiT, Tencent}
  \city{Shenzhen}
  \country{China}
}
\email{xiuqianghe@tencent.com}

\renewcommand{\shortauthors}{Xing Tang, et al.}

\copyrightyear{2024}
\acmYear{2024}
\setcopyright{acmlicensed}\acmConference[RecSys '24]{18th ACM Conference on Recommender Systems}{October 14--18, 2024}{Bari, Italy}
\acmBooktitle{18th ACM Conference on Recommender Systems (RecSys '24), October 14--18, 2024, Bari, Italy}
\acmDOI{10.1145/3640457.3688101}
\acmISBN{979-8-4007-0505-2/24/10}



\begin{abstract}
As user behaviors become complicated on business platforms, online recommendations focus more on how to touch the core conversions, which are highly related to the interests of platforms. These core conversions are usually continuous targets, such as \textit{watch time}, \textit{revenue}, and so on, whose predictions can be enhanced by previous discrete conversion actions. Therefore, multi-task learning (MTL) can be adopted as the paradigm to learn these hybrid targets. However, existing works mainly emphasize investigating the sequential dependence among discrete conversion actions, which neglects the complexity of dependence between discrete conversions and the final continuous conversion. Moreover, simultaneously optimizing hybrid tasks with stronger task dependence will suffer from volatile issues where the core regression task might have a larger influence on other tasks. In this paper, we study the MTL problem with hybrid targets for the first time and propose the model named Hybrid Targets Learning Network (HTLNet) to explore task dependence and enhance optimization. Specifically, we introduce label embedding for each task to explicitly transfer the label information among these tasks, which can effectively explore logical task dependence. We also further design the gradient adjustment regime between the final regression task and other classification tasks to enhance the optimization. Extensive experiments on two offline public datasets and one real-world industrial dataset are conducted to validate the effectiveness of HTLNet. Moreover, online A/B tests on the financial recommender system also show that our model has improved significantly. Our implementation is available here\footnote{\url{https://github.com/fuyuanlyu/HTLNet}}.
\end{abstract}

\maketitle

\section{introduction}

\begin{figure}[htbp]
    \centering
    \includegraphics[width=0.9\linewidth]{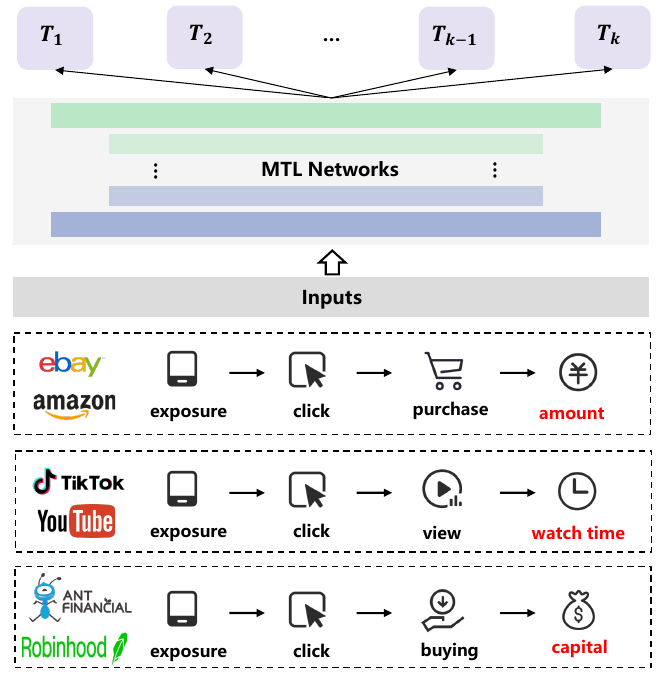}
    \vspace{-10pt}
    \caption{Illustration of hybrid target learning paradigm in various business platforms. The red one denotes core targets.}
    \Description{Illustration of hybrid target learning paradigm in various business platforms. The red one denotes core targets.}
    \label{fig:intro}
\end{figure}

Recommender systems (RS) have played a crucial role in online business platforms, which provide personalized candidates (e.g., ads, videos, funds, apps, etc) based on user interests~\cite{RS}. Usually, the recommendation model in these platforms is required to meet the demand of multiple objectives due to the increasing complexity of user behaviors. For example, the fund recommendation model should consider the likelihood of the user clicking a fund and the likelihood of the user purchasing the fund~\cite{esmm}. However, online platforms have begun to make efforts to touch the core conversion~\cite{revman,click2revenue,watch_time}. As shown in Figure~\ref{fig:intro}, video-sharing platforms value most users' watch time, and fund investment platforms care how much the user invests. Along with preceding discrete click or purchase targets, these continuous core targets consist of hybrid targets. Hence, learning tasks with hybrid targets simultaneously in the recommendation model benefits the online platforms.

Previously, multi-task learning (MTL)~\cite{mtl} has been introduced in the recommendation to make predictions for multiple targets, attracting much attention and making a success in lots of application~\cite{survey,metabalance,pimm,esmm,adatt}. By jointly training multiple tasks in a single model, the MTL can improve performance and decrease computational cost by knowledge sharing and cross-task transferring~\cite{transfer}. 
A vital issue for the success of MTL models is modeling relationships between multiple tasks. Unlike implicit relationships that require a well-designed learning paradigm, the explicit task dependence must be explored among these targets~\cite{AITM, SDMTL}. As an example illustrated in Figure~\ref{fig:intro}, recommending a video with long \textit{watch time} is dependent on the task of whether the user \textit{view} it, which sequentially depends on whether the user \textit{click} it. This prior is greatly helpful for modeling these targets simultaneously in the sequential dependence MTL (SDMTL) paradigm. 

However, the differences between the core and preceding targets make the modeling more challenging. On the one hand, the divergence exists in the modeling objectives. The task of the core target aims to model the distribution of continuous value, compared with preceding tasks modeling the click/conversion rate. Nevertheless, the high conversion rates seldom indicate the high value of core targets, which is contrary to the assumption that a high click rate indicates a high conversion rate in sequential dependence multi-task learning~\cite{esmm,escm2}. Therefore, exploring task dependence among hybrid targets is more complicated than traditional MTL. On the other hand, the tasks of core targets are usually regression tasks, while the preceding tasks are classification tasks. Obviously, the gradients based on regression loss will not be in accordance with the gradients of classification loss from the view of magnitude and direction. This fact will lead to training stability, and thus degrading the performance~\cite{stability}. Hence, the optimization strategy designed for hybrid target learning is necessary.

There have been plenty of state-of-the-art general MTL models for recommendation. Some mainly investigate how to learn implicit task relationships and extract corresponding representations. Usually, they introduce an ensemble of expert submodules and gating networks to learn task relationships. Although many efforts have been devoted to learning implicit relationships among tasks~\cite{mmoe,ple,adatt}, these approaches are still limited in two aspects. First, the negative information will transfer with the shared structure in an unpredictable manner. No matter whether the two tasks are related, the information will pass between them. 
The task dependence priors are absent in modeling, which makes the prediction more difficult. As a result, some methods resort to formulating the learning paradigm as SDMTL~\cite{AITM,SDMTL,esmm,escm2,ctnocvr}. Assuming all the targets are discrete, some specific designed architectures and losses are proposed. However, these works neglect the complexity of the dependence between the classification and regression tasks. As SDMTL assumes $\hat{y}^k-\hat{y}^{k-1}=P(t_k=0,t_{k-1}=1)$, implying the difference of prediction scores between adjacent tasks $(k-1,k)$ is the probability of the task $k$ not happening while the task $k-1$ is observed. However, the equation can no longer hold when the prediction is continuous since we cannot conclude that different continuous values lead to the same probability. Besides, some works focus on optimization strategies to enhance performance. Gradients between tasks should be handled to avoid training instability and performance deterioration ~\cite{metabalance,gradientsurgery,gradnorm,stability}. These methods have no special preference for hybrid target learning, which should seriously consider the fact that regression tasks may have a dominant influence on the network gradients. 

In this paper, we address the hybrid targets learning problem by proposing a Hybrid Targets Learning Net (HTLNet). There are two main challenges identified for our HTLNet. The first challenge is how to explore the dependence among hybrid targets. We tackle this challenge by introducing label embedding and information fusion units in our model. The label embedding unit enables the adjacent tasks to transfer the prediction information following task dependence, and the information fusion unit ensures the information is adaptively transferred. Therefore, HTLNet explores task dependence among hybrid targets through explicit label embedding and implicit task representation.
The second challenge is how to optimize the complex model. Optimization is extremely hard to handle as HTLNet aims to train regression and classification tasks in a unified model. Therefore, we further develop an optimization strategy to conquer the gradients in a shared layer. Our major contributions are summarized as follows:
\begin{itemize}[topsep=0pt,noitemsep,nolistsep,leftmargin=*]
\item This paper first distinguishes the hybrid targets learning problem in the recommendation, which aims to model the sequential auxiliary discrete targets and core continuous targets in an MTL model concerning the core target.
\item We propose a novel model HTLNet that incorporates label and task information to touch the core task. By developing a corresponding optimization strategy, HTLNet can effectively explore the dependence among hybrid targets in a stable way.
\item Extensive offline experiments on public and real-world product datasets are conducted. The results demonstrate the superiority of the proposed model. Furthermore, online experiments also verify the stability and effectiveness of HTLNet.
\end{itemize}

We organize the rest of the paper as follows. In Section 2, we briefly introduce related works. Section 3 formulates the MTL problem with hybrid targets in the recommendation. In section 4, HTLNet is introduced in detail. In Section 5, experimental details are given to verify our HTLNet. Finally, we conclude this paper in Section 6.
\section{Related Work}
Multi-task Learning has broad applications in various fields~\cite{kbs,CVPR,exploring-logically,NMTR,tkdd,optmsm,MTL-A1,MTL-A2,MTL-A3,MTL-A4,MTL-A5}. In recommendation, many MTL network architectures are designed based on Multi-gate Mixture-of-Experts (MMoE)~\cite{mmoe}. Progressive Layered Extraction (PLE)~\cite{ple} separates task-common and task-specific parameters to avoid parameter conflict. AdaTT~\cite{adatt} leverages a residual mechanism and a gating mechanism for task-to-task fusion, which adaptively learns both shared knowledge and specific knowledge. This research line fails to utilize the dependence among targets, which is far from satisfactory in modeling sequential dependence multi-task learning (SDMTL). A typical application of SDMTL is estimating post-click conversion rate (CVR)~\cite{esmm,escm2,dcmt,ctnocvr}. These works formulate the problem as $pCTCVR=pCTR\times pCVR$ and employ an auxiliary click-through rate (CTR) task to enhance the performance. Besides, DCMT~\cite{dcmt} and ESCM$^2$ introduce imputation and inverse propensity weighting tasks to predict unbiased CVR estimation. Despite two dependent tasks, some methods expand the problem into multiple sequential dependence tasks~\cite{AITM,SDMTL,pimm}. AITM~\cite{AITM} proposes a novel information transfer mechanism between different conversion steps to explore the dependence among these targets. TAFE~\cite{SDMTL} investigates the differences between SDMTL and MTL problems and proposes a feature learning module to tackle SDMTL. However, these works assume the targets are discrete conversion steps, neglecting the complexity of hybrid targets learning. Finally, the training stability of multitask ranking models has attracted attention recently~\cite{stability}. Besides, optimization strategy has been a mainstream way to enhance the performance~\cite{metabalance,gradientsurgery, gradnorm}. These works are all based on the general MTL network, which does not consider the network architecture of hybrid targets learning.
\section{preliminary}
This section will elaborate on the MTL problem with hybrid targets in the recommendation. Generally, the recommendation model will rank items according to many objectives including one or more conversion targets and the core target. 

First, consider the prediction of the core target in recommendation over an input space $\mathbf{X}$, where $\mathbf{X}=\{\mathbf{x}_1,\mathbf{x}_2,\cdots,\mathbf{x}_n\}$ represents input features. Given a large dataset of data points $\{\mathbf{x}_i, \mathit{y}_i\}$ where $\mathbf{x}_i \in \mathbf{X} $ denotes the feature vector and $\mathit{y_i}\in\mathbb
{R}$ is the corresponding core continuous target. 
Then, we assume it takes the user $T$ conversion steps to complete the core targets. In each step $t$, $\mathit{y_i^t} \in \{0,1\}$ denotes whether the user completes the conversion and satisfies the constraint in sequential dependence that $y_i^1\geqslant y_i^2\geqslant \cdots \geqslant y_i^T$. Therefore, $\{\mathbf{y}, \mathbf{y}^1, \mathbf{y}^2, \cdots,\mathbf{y}^T\}$ consists of hybrid targets for all the data points. Meanwhile, we introduce these auxiliary sequential tasks to improve the core target's prediction accuracy and provide predictions of these conversions. The MTL problem with hybrid targets is then formulated as follows:

\begin{equation}
\begin{aligned}
&\hat{\mathit{y}}_i = f(\mathbf{x}_i | \Theta, \mathit{y_i^1}, \mathit{y_i^2}, \cdots, \mathit{y_i^T}) \\
&\hat{\mathit{y}}^T_i = f(\mathbf{x}_i | \Theta,\mathit{y_i^1},\mathit{y_i^2},\cdots,\mathit{y_i^{T-1}}) \\
&\cdots \\
&\hat{\mathit{y}}^1_i = f(\mathbf{x}_i | \Theta),
\label{eq:MTL}
\end{aligned}
\end{equation}

where $\Theta$ denotes the parameters of MTL model $f$, $\hat{\mathit{y}}_i$ is the prediction of core target and $\hat{\mathit{y}}^t_i,t\in\{1,\cdots,T\}$ are predictions of preceding targets. 
Note that the first equation denotes the regression task aiming to improve the core target prediction, and the remaining equations are classification tasks, composing a typical sequential dependence MTL (SDMTL)~\cite{SDMTL} respectively. One critical issue for learning this problem is exploring the dependence among hybrid targets with the MTL model according to Equation.\ref{eq:MTL}.

The parameters $\Theta$ of the MTL model usually consist of three components, i.e.
$\{\theta_s, \theta_{core},\{\theta_t\}_{t=1}^T\}$, where $\theta_s$ is the shared parameters, $\theta_{core}$ is the parameters for core targets prediction, and $\theta_t$ denotes task specific parameters. The $\Theta$ is learned by jointly minimizing the core target task loss and other task losses:
\begin{align}
\mathcal{L}(f(\mathbf{X};\Theta), \{\mathbf{y}, \{\mathbf{y}^t\}_{t=1}^T\})= 
\mathcal{L}_{core}(f(\mathbf{X};\theta_s,\theta_
{core}),\mathbf{y}) + \notag\\
\sum_{t=1}^T\mathcal{L}_t(f(\mathbf{X};\theta_s,\theta_t),\mathbf{y}^t)
\end{align}
where $\mathcal{L}_{core}$ can be either mean squared error(MSE)~\cite{watch-time}, and $\mathcal{L}_t$ is the Binary Cross-Entropy (BCE) loss due to the hybrid targets. With one optimizing step, the gradient for the MTL model is computed as follows:
\begin{equation}
G = \nabla_\Theta \mathcal{L}_{core} + \sum_{t=1}^T\nabla_\Theta \mathcal{L}_t
\label{eq:gradient}
\end{equation}
As shown in Equation.\ref{eq:gradient}, one optimization step relies on the gradient, which is composed of two parts: the gradient for the core targets task and the sum of gradients for other tasks.
The gradient component with a larger magnitude can influence the gradient more. Another critical issue for learning the MTL problem with hybrid targets is stabilizing the optimization and enhancing performance~\cite{stability}.
\section{method}

\begin{figure*}[htb]
    \centering
    \includegraphics[width=0.8\linewidth]{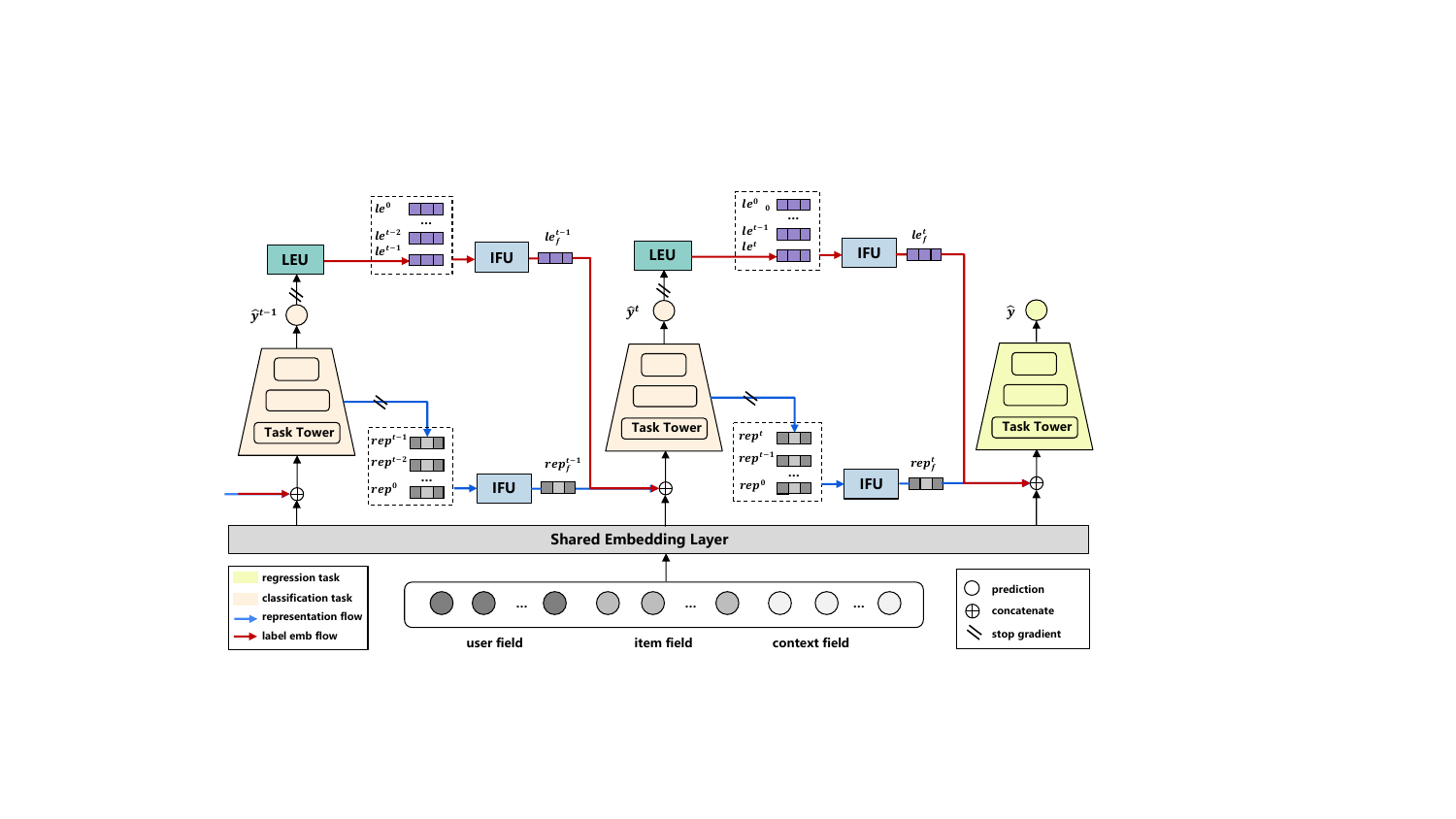}
    \caption{The overall framework of HTLNet. LEU represents the Latent Embedding Unit, which encodes label information, and the Information Fusion Unit, which adaptive infuses all preceding tasks.}
    \Description{The overall framework of HTLNet.}
    \label{fig:framework}
    \vspace{-10pt}
\end{figure*}

To learn the hybrid targets effectively with the MTL model, we propose a novel network architecture, Hybrid Targets Learning Network (HTLNet).  The overall framework of HTLNet is illustrated in Figure~\ref{fig:framework}, which mainly consists of three components: (1) task towers for hybrid targets, (2) the label embedding unit (LEU) encoding the labels of auxiliary sequential tasks, (3) the information fusion unit (IFU) utilizing the information of all preceding tasks. Note that the latter two components make our model explore task dependence effectively. Furthermore, a gradient adaption method based on the HTLNet is also introduced, which enables the optimization to be stable and enhances the model performance. 

\subsection{Framework of HTLNet}

We first introduce the details of HTLNet. The core idea is to explore the task dependence among hybrid targets by introducing label embedding and transferring both implicit and explicit information among tasks.  As shown in Figure~\ref{fig:framework}, all the tasks are shared with an embedding layer. With the input $\mathbf{x}_i$, the features usually include the user, item, and context features in the recommendation, which are denoted as a vector with $m$ fields:
\begin{equation}
\mathbf{x}_i = \{x_{i_1}, x_{i_2}, \cdots, x_{i_m}\}
\end{equation}

Following the input, an embedding matrix $\mathbf{E}\in R^{M\times d}$, where $M$ is the number of features, and $d$ is the embedding dimension, is used in the shared embedding layer to get the corresponding embedding vector for $j-th$ field:
\begin{equation}
e_{i_j} = x_{i_j}\mathbf{E}.
\end{equation}
Then, the vectors are all stacked together as shared embedding vector, $\mathbf{e_i}=[e_{i_1},e_{i_2},\cdots,e_{i_m}]$. 

Notice that the core continuous target introduces the difficulty of sharing information between hybrid targets. To explore the task dependency relationship, it is not enough to depend solely on shared embedding. Moreover, the prediction of the core target will benefit from the information from the preceding tasks.
Therefore, we incorporated explicit and implicit information transferred from preceding tasks as input to the task tower. Specifically, for two adjacent tasks $t-1$ and $t$, the input for task $t$ is computed as:
\begin{equation}
h^t_{input,i} = concat(rep_f^{t-1}, le_f^{t-1}, \mathbf{e_i}),
\end{equation}
where $rep_f^{t-1}$ and $le_f^{t-1}$ are the outputs from the IFUs of task $t-1$ respectively. 

Notice that the $le_f^{t-1}$ is the explicitly label information, and the $rep_f^{t-1}$ is the implicit information from the task tower, which can be the output from a particular layer of the task tower. One way to encode label information is to use it directly as a constraint in the final loss~\cite{AITM,SDMTL}, which is inappropriate in hybrid target learning and will make the prediction of the core target suffer from the deficiency of preceding label information. We introduce the LEU here to address the issue, which will be illustrated in detail afterward. 

As to the classification tasks, the task tower gives the prediction probability of instance $i$:
\begin{equation}
\hat{y}_i^{t} = \sigma(MLP(h_{input,i}^t)),
\end{equation}
where the MLP is multi-layer perception, and $\sigma(x)=\frac{1}{1+e^{-x}}$ is the sigmoid function. The loss for these tasks is usually binary cross entropy:
\begin{equation}
\mathcal{L}_{t} = -\frac{1}{n}\sum_{i=1}^n [y_i^t \text{log}(\hat{y}_i^t) + (1-y_i^t) \text{log}(1-\hat{y}_i^t)]
\end{equation}
As to the core continuous target, the task tower gives the numerical prediction of instance $i$:
\begin{equation}
\hat{y}_i = MLP(h_{input,i}),
\end{equation}
and the loss can be MSE:
\begin{equation}
\mathcal{L}_{core}= \frac{1}{n}\sum_{i=1}^{n}(y_i-\hat{y}_i)^2 
\end{equation}

\subsubsection{Label Embedding Unit}



\begin{figure}
\centering
\subfigure[Label Embedding Unit]{
\begin{minipage}[t]{0.45\linewidth}
\centering
\includegraphics[width=\textwidth]{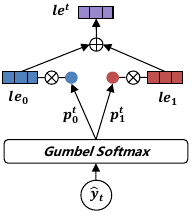}
\label{fig:leu}
\end{minipage}
}
\subfigure[Information Fusion Unit]{
\begin{minipage}[t]{0.45\linewidth}
\centering
\includegraphics[width=\textwidth]{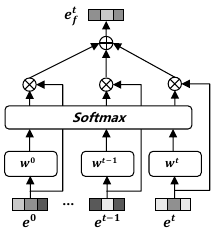}
\label{fig:ifu}
\end{minipage}
}
\vspace{-10pt}
\caption{Illustration of Label Embedding Unit and Information Fusion Unit. $\otimes$ denotes element-wise multiplication. Note that $p^t_0+p^t_1=1$.}
\Description{Illustration of Label Embedding Unit and Information Fusion Unit.}
\label{fig:leu-ifu}
\vspace{-10pt}
\end{figure}

The LEU aims to encode the label information explicitly, as shown in Figure~\ref{fig:leu}. Unlike the previous study~\cite{exploring-logically}, which solely focused on implicit information among task towers, encoding label information can let a task utilize the labels of all its preceding tasks as formulated in Equation~\ref{eq:MTL}. However, the discrete $0/1$ label information can hardly be delivered directly to the core continuous target due to the hybrid targets.

Therefore, an embedding table consisting of two rows is introduced for each classification task in our setting, which is $E^t\in R^{2\times L_d}$ for label $0/1$. The embedding table enables a one-to-one mapping from a label to a $L_d$-dimensional trainable vector. Each row contains the information corresponding to the label $0$ or $1$, denoted $le_0$ and $le_1$, respectively.
An alternative to mapping the embedding is using the golden labels of tasks, which will introduce the train-test discrepancy. Specifically, the golden labels are only available in the training stage, which is missing in the test set. Moreover, directly using prediction labels for tasks leads to unstable training due to cascading errors between tasks. 

In our LEU, we sample a label from the predicted probability distribution instead of predefined labels. If the sampling label is a misleading prediction of the final core task, the back-propagated gradients from the core target will penalize the embedding. The only issue is the sampling operation is not differentiable, which is incompatible with our framework.
To address this issue, we further introduce the Gumbel-softmax re-parameter operation~\cite{gumbel} here to approximate the sampling by:
\begin{align}
p^t = \frac{exp(\frac{log(\hat{y}^t)}{\tau})}{exp(\frac{log(\hat{y}^t)}{\tau})+exp(\frac{log(1-\hat{y}^t)}{\tau})},
\end{align}
where $\tau$ is the decayed temperature parameter to control the smoothness of the Gumbel-softmax, when $\tau$ approximates to zero, the output will be a two-dimensional one-hot vector. The Gumbel-softmax allows HTLNet to calculate the soft weighted embedding value instead of hard selecting particular embedding, thus significantly reducing cascading errors. Its detailed motivation is further discussed in Appendix~\ref{ap:gumbel}. The output of LEU is then formalized as the weighted sum over the label embedding table:
\begin{equation}
le^t = [p^t, 1-p^t] E^t.
\end{equation}
Unlike directly utilizing the discrete label, this label embedding mechanism can transfer label information from discrete targets to the core continuous target. 

\subsubsection{Information Fusion Unit}
The IFU is proposed to fuse all the information from the preceding tasks. Combined with LEU, there are two kinds of information to fuse, i.e., label embedding and task representation. The IFU adopts a similar attention mechanism in~\cite{AITM,attention} as illustrated in Figure~\ref{fig:ifu}.

 
Specifically, the preceding task vectors for task $t$ can be denoted as $[\mathbf{e}^0, \mathbf{e}^1, \cdots, \mathbf{e}^t]$, where $\mathbf{e}$ is either $le$ from LEU or $rep$ from task tower representation. The attention is designed to allocate the weights of these transferred information adaptively.

\begin{equation}
\mathbf{e}_f^t=\sum_{u\in [\mathbf{e}^0, \mathbf{e}^1, \cdots, \mathbf{e}^t]}w^t h_1(\mathbf{u})
\end{equation}
where $w^t$ is formulated as:
\begin{align}
w^t=\frac{exp(\hat{w}_\mathbf{u})}{\Sigma_u exp(\hat{w}_\mathbf{u})},\quad
\hat{w}_\mathbf{u} = \frac{<h_2(\mathbf{u}), h_3(\mathbf{u})>}{\sqrt{k}}
\end{align}
Where $<\centerdot,\centerdot>$ denotes the dot product, $k$ is the hidden dimension, and $h_1(\centerdot),h_2(\centerdot),h_3(\centerdot)$ are learnable kernels for transforming input information into a new output space. Finally, we get label information $le^t_f$ and task representation $rep^t_f$ from IFU to explore the task dependence effectively.

\subsection{Optimization Strategy for HTLNet}
\label{sec:opt}

Despite exploring the task dependence, the designed network poses some optimization challenges. First, the loss of core continuous targets differs from other tasks, which means the gradients from the core target tower are very far from other tasks, leading to performance deterioration. Second, all the label embeddings and task representations will be transferred to the core target tower, allowing the core target loss to mainly influence the optimization of other tasks with the gradient backpropagation to label prediction and task representation. Hence, we are motivated to propose an optimization strategy for our HTLNet.

Two ways for a task to influence other tasks are shared embedding and transferred information. However, as discussed above, 
the transferred information will disturb the preceding task's label prediction and task representation by LEU and IFU. Hence, we cut off the influence of transferred information, leaving the shared embedding as the only way to influence each other:
\begin{align}
\label{eq:stop}
\text{LEU}(\mathit{stop\_gradient}(\hat{y}^t)),\quad \text{IFU}(\mathit{stop\_gradient(rep^t)}).
\end{align}

Then, we target dealing with the gradient of shared embedding from the view of direction and magnitude. Denoted $G_{core}=\nabla_\Theta\mathcal{L}_{core}$ and $G_t=\nabla_\Theta\mathcal{L}_{t}$ in Equation~\ref{eq:gradient}, we consider the pairs of task gradients $\{G_{core}, G_t\}_{t=1}^T$ to cope with the gradient conflict of shared embedding between discrete tasks and our core continuous task. For every classification task $t$ and core task, we first eliminate the gradient direction conflict with the core task being the  target gradient for shared embedding~\cite{gradientsurgery}:
\begin{align}
&G_t = G_t-\alpha \frac{G_t\bullet G_{core}}{\lVert G_{core} \rVert^2}G_{core}\notag\\ 
&s.t.\quad G_t\bullet G_{core} < 0,
\label{eq:direction}
\end{align}
where $\alpha$ is a hyperparameter to control the amount of removing the conflict. As a variant of method in~\cite{gradientsurgery}, when the gradient of preceding task $t$ conflicts with the core task, meaning cosine similarity is negative, the Equation~\ref{eq:direction} projects the $G_t$ onto the normal plane of $G_{core}$. After iterations, all the $G_t$s will not conflict with $G_{core}$, which means all preceding tasks are optimized towards core tasks.

Besides gradient conflict, another issue is the magnitude discrepancy between $G_{core}$ and $G_t$. Notice that $\mathcal{L}_{core}$ is MSE loss, the magnitude of which is usually more significant than the Logloss for $\mathcal{L}_t$.
Inspired by ~\cite{metabalance}, we adaptive adjust the gradient magnitude of task $t$ toward the core task in a direct way:
\begin{align}
G_t = \gamma\cdot \frac{\lVert G_{core} \rVert}{\lVert G_{t} \rVert} \cdot G_t + (1-\gamma)\cdot G_t, 
\label{eq:magnitude}
\end{align}
where $\gamma$ is a hyperparameter tuned to balance the gradient magnitude between task $t$ and the core task,$\frac{\lVert G_{core} \rVert}{\lVert G_{t} \rVert}$ is the calculated weight to adjust the gradients. We introduce a hyperparameter $C$ to clip the weight to avoid the explosion of gradients in a batch. When $\frac{\lVert G_{core} \rVert}{\lVert G_{t} \rVert}$ is too large, it will clip up to $C$, while it is too small, it will clip down to $\frac{1}{C}$. Combining both direction and magnitude, we summarize the gradient process for shared embedding in the Algorithm~\ref{alg:opt}.

\begin{algorithm}
\caption{Gradient for shared parameters}
\label{alg:opt}

\KwIn{Tasks shared model parameters $\theta_s$, core target task loss $\mathcal{L}_{core}(\theta_s)$, preceding sequential task losses $\mathcal{L}_{t}(\theta)$, projection relax factor $\alpha$, balance relax factor $\gamma$, balance clip threshold $\mathcal{C}$}

\KwResult{Gradient of shared model parameters $G_s$}
$G_{core}(\theta_s) \leftarrow \nabla_{\theta_s} \mathcal{L}_{core}(\theta_s)\;$

\For{$t\leftarrow 1\;to\;T$ }{
    $G_{t}(\theta_s) \leftarrow \nabla_{\theta_s} \mathcal{L}_{t}(\theta_s)$\
    
    \If{$G_{t} \cdot G_{core}<0$}
        {$G_{t} =G_{t}-\alpha\frac{G_{t} \cdot G_{core}}{\lVert G_{core}\rVert^{2}} G_{core}\;$}
    
    $weight$ = $\frac{\lVert G_{core}\rVert}{\lVert G_{t}\rVert}$\;
    
    \If{$weight<1/\mathcal{C}$}
        {$weight\leftarrow 1/\mathcal{C}$}
    \If{$weight>\mathcal{C}$}
        {$weight\leftarrow \mathcal{C}$}
    {$G_{t}=\gamma \cdot weight\cdot G_{t} + (1-\gamma) \cdot G_{t}$}
   } 
$G_{s} = G_{core} + \sum_{t=1}^{T} G_{t}$ \BlankLine 
$\textbf{ouput} \;  G_{s}$
\end{algorithm}

\section{experiments}

In this section, we first outline the experimental setup for our model. There are two public and one private industry dataset to evaluate the performance. We design experiments to answer the following research questions:
\begin{itemize}[topsep=0pt,noitemsep,nolistsep,leftmargin=*]
\item \textbf{RQ1}: Does HTLNet achieve superior performance on MTL for hybrid targets compared with mainstream MTL models?
\item \textbf{RQ2}: How do the proposed network structure and optimization strategy affect the HTLNet? Can we only rely on one component to achieve the same performance?
\item \textbf{RQ3}: How well do the proposed LEU and IFU affect the performance of HTLNet?
\item \textbf{RQ4}: Does the optimization strategy for HTLNet perform better than the previous methods?
\item \textbf{RQ5}: Can the HTLNet also achieve better performance in the online setting?
\end{itemize}

\subsection{Experimental Setup}

\subsubsection{Datasets}
The datasets in previous MTL works only consist of tasks with a single type of label, which can not be directly used for evaluation in hybrid target learning. Hence, we define related tasks with hybrid targets from two public datasets, including video and revenue scenarios, and collect one product dataset from a real-world fund recommendation scenario.
\begin{itemize}[leftmargin=*]
\item KuaiRand~\cite{kuairand}: This is a sequential recommendation dataset collected from the video-sharing mobile app. Despite providing randomly exposed items, it also contains various user behaviors, such as clicking, sharing, and liking the app. In our experiments, we use the KuaiRand-pure and choose three hybrid targets defined in the dataset: click, long view, and watch time, setting watch time as core target. 
\item Kaggle-Revenue~\cite{ziln}: This Kaggle Acquire Valued Shoppers Challenge competition dataset contains a complete basket-level shopping history for customers and companies. The data process follows~\cite{ziln}, and we define three hybrid targets: user repurchase in one year, user repurchase in one month, and the amount of user repurchase in one month, which is the core target.
\item Product: The product dataset is collected from the recommendation logs from an online fund recommendation platform in the past three months, which records users' click and purchase fund behaviors. We define three hybrid targets: click, purchase, and the amount of purchase.
\end{itemize}

The statistics of these three datasets are shown in Table~~\ref{tab:datasets}. The datasets are divided chronologically into training, validation, and test sets, keeping the ratio $8:1:1$.


 

  

   


\begin{table}[!htbp]
\caption{Statistics of the two public datasets and one product dataset.}
\vspace{-10pt}
\centering
\resizebox{\linewidth}{!}{
\begin{tabular}{c|c|cc|cc}
\hline
\multicolumn{6}{c}{KuaiRand} \\
\hline
name & sample & \multicolumn{2}{c|}{watch time} & click & long view \\
\hline
metrics & size & mean & std & ratio & ratio \\
value & 1186059 & 6.94 & 18.79 & 17.62\% & 8.50\% \\        
\hline
\hline
\multicolumn{6}{c}{Kaggle-Revenue} \\
\hline
name & sample & \multicolumn{2}{c|}{repurchase 1M amt} & repurchase 1Y & repurchase 1M \\
\hline
metrics & size & mean & std & ratio & ratio \\
value & 5041007 & 4.52 & 165.69 & 85.68\% & 41.05\% \\
\hline
\hline
\multicolumn{6}{c}{Product}   \\
\hline
name & sample & \multicolumn{2}{c|}{purchase amt} & click & purchase \\
\hline
metrics & size & mean & std & ratio & ratio \\
value & 19160294 & 1627.05 & 8009.42 & 8.65\% & 4.82\% \\
\hline
\end{tabular}}
\label{tab:datasets}
\end{table}

\subsubsection{Baselines}
We choose baseline models from four aspects. First, the \textbf{DNN} is a three-layer MLP structure for a single task. Second, \textbf{Shared-Bottom}~\cite{mtl}, \textbf{MMoE}~\cite{mmoe}, \textbf{PLE}~\cite{ple}, and \textbf{AdaTT}~\cite{adatt} are SOTA models in MTL for recommendation. Third, \textbf{ESMM}~\cite{esmm} and \textbf{AITM}~\cite{AITM} are MTL models that take the sequential dependence into consideration. Finally, \textbf{MetaBalance}~\cite{metabalance} is the recent optimization method for MTL. The implementation details are discussed in Appendix~\ref{ap:offline}.

\subsubsection{Evaluation Metrics}
Because our experiments have hybrid targets, we introduce the evaluation metrics for the classification and regression tasks, respectively. Two widely used metrics, AUC and LogLoss, are adopted for all the classification tasks. 
Regarding regression tasks, we adopt NRMSE (Normalized Root-Mean-Square Error) and NMAE (Normalized Mean Absolute Error) to evaluate watch time and purchase amount following~\cite{ltv,watch-time}. Besides these two metrics for regression, Gini and Spearman~\cite{ziln} are also adopted to measure purchase amounts because this prediction aims to rank the users according to purchase amounts.
Note that AUC, Gini, and Spearman all measure the rank results; a more significant value indicates a better performance. On the contrary, Logloss, NRMSE, and NMAE suggest a smaller value will lead to a better result. Detailed definition of these metrics are listed in Appendix~\ref{ap:metric}.

\begin{table*}[!htbp]
\caption{Results on two public and one product datasets.}
\vspace{-10pt}
\label{tab:main_result_both}
\centering
\begin{tabular}{c|c|c|cccccccc|c}
\hline
    Dataset & Task & Metrics & DNN & SB & MMoE & PLE & AdaTT & ESMM & AITM & MB & HTLNet \\
\hline
    \multirow{6}{*}{KuaiRand} 
    & \multirow{2}{*}{click} 
    & AUC$\uparrow$ & 0.7512 & 0.7399 & 0.7379 & 0.7409 & 0.7396 & 0.7303 & 0.7391 & \underline{0.7541} & \textbf{0.7574}$^*$ \\
    & & Logloss$\downarrow$ & 0.4074 & 0.4115 & 0.4127 & 0.4110 & 0.4120 & 0.4259 & 0.4118 & \underline{0.4065} & \textbf{0.4035}$^*$ \\
\cline{2-12}
    & \multirow{2}{*}{long view} 
    & AUC$\uparrow$ & 0.7698 & 0.7685 & 0.7645 & 0.7694 & 0.7683 & 0.7540 & 0.7672 & \underline{0.7768} & \textbf{0.7787}$^*$ \\
    & & Logloss$\downarrow$ & 0.2536 & 0.2543 & 0.2566 & 0.2542 & 0.2550 & 0.2832 & 0.2552 & \underline{0.2527} & \textbf{0.2509} \\
\cline{2-12}
    & \multirow{2}{*}{watch time}
    & NRMSE$\downarrow$ & 0.8979 & 0.8968 & 0.8970 & \underline{0.8963} & 0.8975 & 0.8981 & 0.8978 & 0.8967 & \textbf{0.8937}$^*$ \\
    & & NMAE$\downarrow$ & 0.9553 & 0.9620 & 0.9621 & 0.9648 & \underline{0.9508} & 0.9605 & 0.9533 & 0.9509 & \textbf{0.9093}$^*$ \\
\hline
    \multirow{7}{*}{Kaggle-}
    & \multirow{2}{*}{repurchase 1Y} 
    & AUC$\uparrow$ & 0.6892 & 0.6568 & 0.6696 & 0.6732 & 0.6720 & 0.6361 & 0.6728 & \underline{0.6896} & \textbf{0.6960}$^*$ \\
    \multirow{7}{*}{Revenue} & & Logloss$\downarrow$ & 0.3830 & 0.3931 & 0.3903 & 0.3885 & 0.3888 & 0.6871 & 0.3871 & \underline{0.3825} & \textbf{0.3790}$^*$ \\
\cline{2-12}
    & \multirow{2}{*}{repurchase 1M} 
    & AUC$\uparrow$ & 0.6255 & 0.6134 & 0.6161 & 0.6178 & 0.6187 & 0.6046 & 0.6151 & \underline{0.6267} & \textbf{0.6296}$^*$ \\
    & & Logloss$\downarrow$ & 0.6542 & 0.6606 & 0.6593 & 0.6598 & 0.6580 & 0.7755 & 0.6579 & \underline{0.6539} & \textbf{0.6520} \\
\cline{2-12}
    & \multirow{3}{*}{repurchase 1M} 
    & NRMSE$\downarrow$ & 0.9985 & 0.9984 & 0.9987 & 0.9988 & 0.9987 & 0.9989 & \underline{0.9982} & 0.9985 & \textbf{0.9980} \\
    & \multirow{3}{*}{amount}
    & NMAE$\downarrow$ & 1.0902 & 1.1064 & \underline{1.0623} & 1.0708 & 1.0969 & 1.1981 & 1.1274 & 1.1265 & \textbf{1.0585}$^*$ \\
    & & Spearman$\uparrow$ & 0.2339 & 0.2340 & 0.2357 & 0.2332 & \underline{0.2419} & 0.2379 & 0.2204 & 0.2376 & \textbf{0.2475}$^*$ \\
    & & Gini$\uparrow$ & 0.5093 & 0.5090 & 0.5093 & 0.5136 & \underline{0.5185} & 0.5131 & 0.5029 & 0.5100 & \textbf{0.5233}$^*$ \\
\hline
    \multirow{8}{*}{Product}
    & \multirow{2}{*}{click} 
    & AUC$\uparrow$ & 0.7914 & 0.7968 & 0.7749 & 0.7936 & 0.7958 & 0.7069 & \underline{0.8036} & 0.7965 & \textbf{0.8049}$^*$ \\
    & & Logloss$\downarrow$ & 0.1508 & 0.1498 & 0.1544 & 0.1500 & 0.1497 & 0.2324 & \underline{0.1479} & 0.1492 & \textbf{0.1472} \\
\cline{2-12}
    & \multirow{2}{*}{convert} 
    & AUC$\uparrow$ & 0.8929 & 0.8991 & 0.8770 & 0.8983 & 0.9000 & 0.8378 & 0.9003 & \underline{0.9008} & \textbf{0.9023} \\
    & & Logloss$\downarrow$ & 0.1105 & 0.1062 & 0.1119 & 0.1054 & 0.1040 & 0.1982 & \underline{0.1034} & 0.1035 & \textbf{0.1030} \\
\cline{2-12}
    & \multirow{3}{*}{purchase} 
    & NRMSE$\downarrow$ & 0.8475 & 0.8441 & \underline{0.8351} & 0.8507 & 0.8401 & 0.8594 & 0.8377 & 0.8365 & \textbf{0.8346} \\
    & \multirow{3}{*}{amount}
    & NMAE$\downarrow$ & 0.8471 & 0.8499 & 0.8846 & 0.9276 & 0.9234 & \underline{0.8469} & 0.8807 & 0.9094 & \textbf{0.8383}$^*$ \\
    & & Spearman$\uparrow$ & 0.1759 & 0.1779 & 0.2060 & \underline{0.2088} & 0.2014 & 0.1892 & 0.1956 & 0.1954 & \textbf{0.2244}$^*$ \\
    & & Gini$\uparrow$ & 0.7982 & 0.8085 & 0.8254 & 0.8274 & \underline{0.8279} & 0.8128 & 0.8212 & 0.8219 & \textbf{0.8362}$^*$ \\
\hline
\end{tabular}
\begin{tablenotes}\footnotesize
\item[*] Here the best and second best results are marked in \textbf{bold} and \underline{underline} fonts, respectively. Each experiment is repeated 10 times for statistical confidence, and $^{*}$ indicates a significance level of $p\leq 0.05$ based on a two-sample t-test between our method and the best baseline.
\end{tablenotes}
\vspace{-5pt}
\end{table*}
\subsection{Overall Performance (RQ1)}


The compared experimental results on two public datasets and one product dataset are separately presented in Table~\ref{tab:main_result_both} due to the space. Overall, our MTLNet achieves the best metrics in all target prediction tasks for all the datasets, which verifies the effectiveness of HTLNet. We summarize our insights as follows.

Firstly, compared with single-task DNN models, most MTL models perform better on all tasks in the corresponding dataset, demonstrating that MTL can predict all hybrid tasks well. This justifies the reason we introduce MTL to learning hybrid targets. Among all the models, our HTLNet can improve performance significantly due to the carefully designed network architecture and corresponding optimization strategy.

Secondly, HTLNet is statistically significant on most tasks compared to the other baselines. This necessitates considering both network architecture design and optimization strategy to improve the performance of hybrid target learning. Specifically, HTLNet performs consistently well on core target tasks and preceding sequential tasks compared with ESMM, AITM, and MetaBalance, which indicates that our HTLNet architecture can explore the dependence between hybrid targets. As to MMoE, AdaTT, and PLE, they achieve better performance on core target tasks compared to others. However, because of the optimization difficulty for hybrid targets, they perform worse on the preceding classification tasks.

Thirdly, among MTL models except for HTLNet, MetaBalance performs better in auxiliary tasks, demonstrating that optimization strategy is a critical factor in hybrid target learning. On the other hand, the general MTL model performs slightly better than the SDMTL model on core target regression tasks, i.e., AITM and ESSM. We conjecture that these SDMTL models are mainly designed to explore the dependence among classification targets, neglecting the peculiarity of regression tasks. On the contrary, our HTLNet first carefully designs the network architecture to explore the dependence among hybrid targets and provide a corresponding optimization strategy, further improving performance.

Moreover, an alternative loss to Kaggle-Revenue and product datasets is lognormal loss~\cite{ziln}, which has been widely adopted in Lifetime Value Prediction. For the experimental soundness, we refer results in Appendix~\ref{ap:lognormal}

\subsection{Ablation Study (RQ2)}

We conduct ablation studies on the KuaiRand dataset to analyze the effectiveness of the main components.

\begin{table*}[!htbp]
\caption{Ablation Study of HTLNet on KuaiRand dataset}
\label{tab:ablation}
\vspace{-10pt}
\begin{tabular}{c|cc|cc|cc}
\hline
 {Dataset} & \multicolumn{6}{c}{KuaiRand} \\
\hline
    {Task} & \multicolumn{2}{c|}{click} & \multicolumn{2}{c|}{long view} & \multicolumn{2}{c}{watch time} \\
\hline
    Metrics & AUC$\uparrow$ & Logloss$\downarrow$ & AUC$\uparrow$ & Logloss$\downarrow$ & NRMSE$\downarrow$ & NMAE$\downarrow$ \\      
\hline
    HTLNet & \textbf{0.7574} & \textbf{0.4035}$^*$ & \textbf{0.7787}$^*$ & \textbf{0.2509} & \textbf{0.8937} & \textbf{0.9093}$^*$ \\ 
    w/o Architecture & 0.7470 & 0.4082 & 0.7746 & 0.2527 & 0.8955 & 0.9630 \\ 
    w/o Optimization & 0.7420 & 0.4104 & 0.7713 & 0.2538 & 0.8970 & 0.9480 \\ 
\hline
\end{tabular}
\begin{tablenotes}\footnotesize \centering
\item[*] Here $^{*}$ indicates a significance level of $p\leq 0.05$ based on a two-sample t-test between HTLNet and the best-performed baseline.
\end{tablenotes}
\vspace{-5pt}
\end{table*}

As discussed in the previous section, our HTLNet contributes to hybrid target learning in network architecture and optimization strategy. Hence, we introduce two variants of HTLNet: (1) \textbf{HTLNet w/o Architecture} replaces the designed architecture with a shared-bottom network, which means there are no special modules for exploring the task dependencies. (2) \textbf{HTLNet w/o Optimization} removes the gradient process described in Algorithm~\ref{alg:opt}, which only relies on the designed architecture.

We present the results in Table~\ref{tab:ablation} with the evaluation metrics on all tasks. It can be concluded that HTLNet outperforms all variants consistently. Specifically, \textbf{HTLNet w/o Optimization} is the worst variant in this setting. This phenomenon further justifies our contribution to propose an optimization strategy based on the designed architecture. This is necessary in hybrid target learning, where the objects vary with each other significantly. Moreover, the \textbf{HTLNet w/o Architecture} is slightly better than \textbf{HTLNet w/o Optimization} while still worse than HTLNet. This indicates that our LEU and IFU effectively transfer information from task labels and representations.

\subsection{Analysis of Network Architecture (RQ3)}

To better understand the effects of architecture in our HTLNet, we further design experiments to investigate the network architecture on the KuaiRand dataset.

\begin{table*}[!htbp]
\caption{Network Architecture Analysis on KuaiRand dataset}
\label{tab:Architecture}
\vspace{-10pt}
\begin{tabular}{c|cc|cc|cc}
\hline
    Dataset & \multicolumn{6}{c}{KuaiRand} \\
\hline
    Task & \multicolumn{2}{c|}{click} & \multicolumn{2}{c|}{long view} & \multicolumn{2}{c}{watch time} \\
\hline
    Metrics & AUC$\uparrow$ & Logloss$\downarrow$ & AUC$\uparrow$ & Logloss$\downarrow$ & NRMSE$\downarrow$ & NMAE$\downarrow$ \\    
\hline
    HTLNet & \textbf{0.7574} & \textbf{0.4035} & \textbf{0.7787} & \textbf{0.2509} & \textbf{0.8937} & \textbf{0.9093}$^*$ \\ 
    w/o representation & 0.7526 & 0.4058 & 0.7773 & 0.2512 & 0.8951 & 0.9248 \\ 
    w/o label embedding & 0.7565 & 0.4044 & 0.7785 & 0.2511 & 0.8950 & 0.9198 \\ 
\hline
\end{tabular}
\begin{tablenotes}\footnotesize \centering
\item[*] Here $^{*}$ indicates a significance level of $p\leq 0.05$ based on a two-sample t-test between HTLNet and the best-performed baseline.
\end{tablenotes}
\vspace{-5pt}
\end{table*}

Notice that two kinds of information are transferred to explore the dependence among hybrid targets. We thus investigate how these two kinds of information affect the performance. We introduce two variants of the network architecture of HTLNet. (1) \textbf{HTLNet w/o representation} removes the IFU for task representations fusion. (2) \textbf{HTLNet w/o label embedding} removes the LEU for each task, equivalent to removing the label information transferred among tasks. The results are summarized in Table~\ref{tab:Architecture}. Overall, removing any information from the original HTLNet will degrade the performance. \textbf{HTLNet w/o representation} performs slightly better than \textbf{HTLNet w/o label embedding}, which indicates the information is limited from preceding tasks with only label embedding.

\begin{figure}[htbp]
    \centering
    \includegraphics[width=0.9\linewidth]{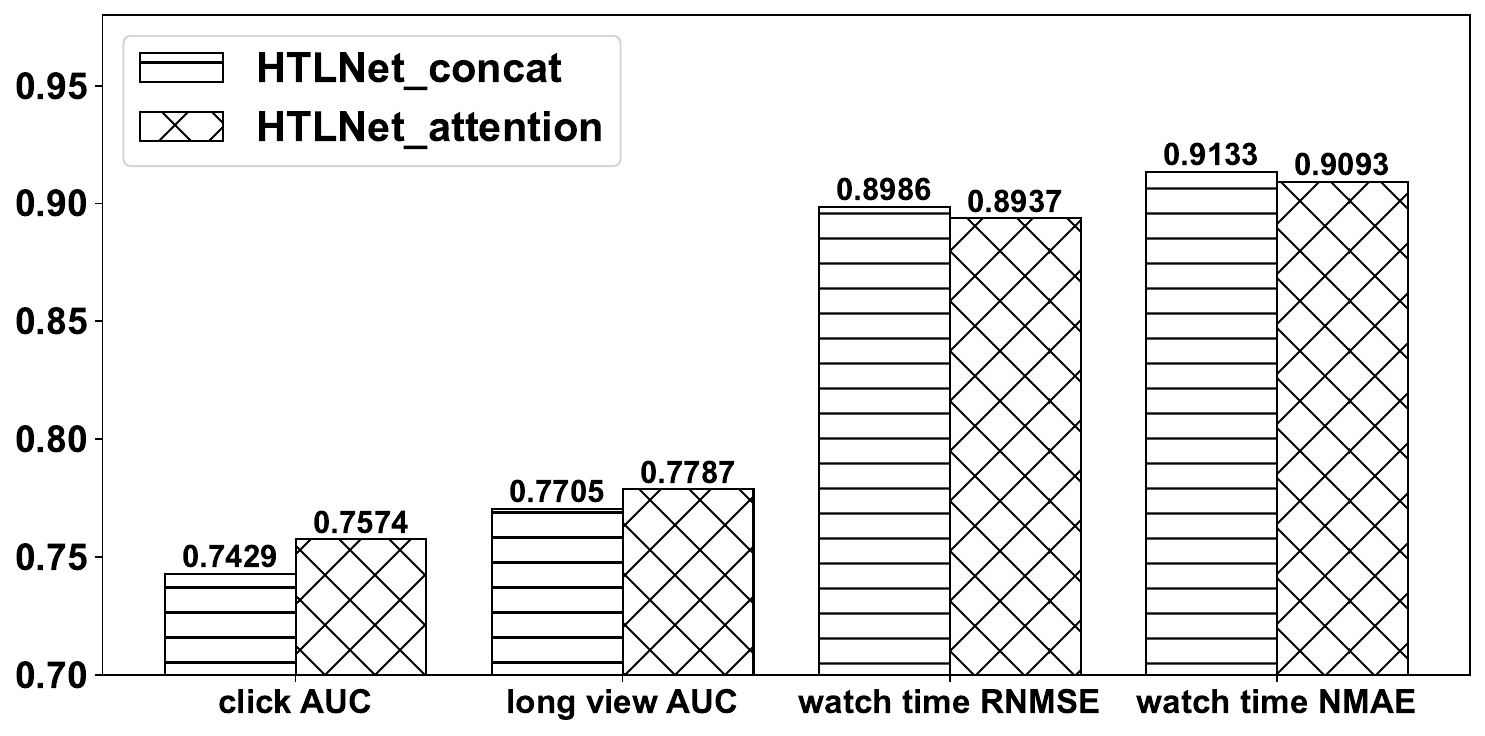}
    \vspace{-10pt}
    \caption{Comparison of the different operations in IFU.}
    \Description{Comparison of the different operations in IFU.}
    \label{fig:operation}
    \vspace{-10pt}
\end{figure}

\subsection{Analysis of Optimization Strategy (RQ4)}

We also investigate the effects of information fusion in HTLNet. As the information fusion operation in IFU is attention, we replace it with a simple concatenation operation. The results are given in Figure~\ref{fig:operation}. 
The performance of classification tasks suffers a drop with the concatenation operation, as directly fusing information from a preceding task cannot avoid the negative transfer. Over the click AUC, HTLNet also performs better with the attention operation, indicating that transferring inappropriate information will influence preceding tasks with shared embedding. As to the core target, the attention mechanism can adaptively learn what and how much information to transfer from preceding tasks. 

\begin{table*}[!htbp]
\centering
\caption{Optimization Strategy Analysis on KuaiRand dataset}
\label{tab:Optimization}
\vspace{-10pt}
\begin{tabular}{c|cc|cc|cc}
\hline
    Dataset & \multicolumn{6}{c}{KuaiRand} \\
\hline
    Task & \multicolumn{2}{c|}{click} & \multicolumn{2}{c|}{long view} & \multicolumn{2}{c}{watch time} \\
\hline
    Metrics & AUC$\uparrow$ & Logloss$\downarrow$ & AUC$\uparrow$ & Logloss$\downarrow$ & NRMSE$\downarrow$ & NMAE$\downarrow$ \\      
\hline
    HTLNet & \textbf{0.7574} & \textbf{0.4035} & \textbf{0.7787}$^*$ & \textbf{0.2509} & \textbf{0.8937} & \textbf{0.9093}$^*$ \\ 
    w/o stop\_gradient & 0.7292 & 0.4214 & 0.7549 & 0.2684 & 0.8974 & 0.9344 \\ 
    w/o gradient conflict & 0.7516 & 0.4067 & 0.7743 & 0.2523 & 0.8946 & 0.9179 \\
    w/o gradient magnitude & 0.7424 & 0.4109 & 0.7710&0.2547&0.8987&0.9321\\
    HTLNet-Gradient Surgery & 0.7432 & 0.4104 & 0.7718 & 0.2546 & 0.8992 & 0.9444 \\ 
    HTLNet-MetaBalance & 0.7517 & 0.4070 & 0.7773 & 0.2519 & 0.8948 & 0.9136 \\ 
\hline
\end{tabular}
\begin{tablenotes}\footnotesize
\centering
\item[*] Here $^{*}$ indicates a significance level of $p\leq 0.05$ based on a two-sample t-test between HTLNet and the best-performed baseline.
\end{tablenotes}
\vspace{-5pt}
\end{table*}

Another critical component of the HTLNet is the optimization strategy. We analyze optimization strategy using several experiments on the KuaiRand dataset. 

The first one compares HTLNet with relevant variants. (1) \textbf{HTLNet w/o stop\_gradient}, \textbf{HTLNet w/o gradient conflict} and \textbf{HTLNet w/o gradient magnitude} represents the HTLNet without any stop gradient operations in Equation.\ref{eq:stop}, HTLNet without gradient direction conflict in Equation.\ref{eq:direction} and HTLNet without gradient magnitude in Equation.\ref{eq:magnitude} respectively. (2) \textbf{HTLNet-Gradient Surgery} employs the Gradient Surgery~\cite{gradientsurgery} method on the HTLNet architecture. (3) \textbf{HTLNet-MetaBalance} employs the MetaBalance method on the HTLNet architecture. The results are illustrated in Table~\ref{tab:Optimization}. From the results, we conclude that the operation of the stop gradient affects the performance significantly among all the gradient operations. We believe that the regression targets interfering with other tasks directly make the optimization volatile. Gradient Surgery and MetaBalance are the optimization methods for general MTL models. The results show that they perform worse than our optimization strategy, which suggests that our optimization method is suitable and customized for hybrid target learning.

\begin{figure}[htbp]
	\centering
	\subfigure[HTLNet without Gradient Process]{\includegraphics[width=0.75\columnwidth]{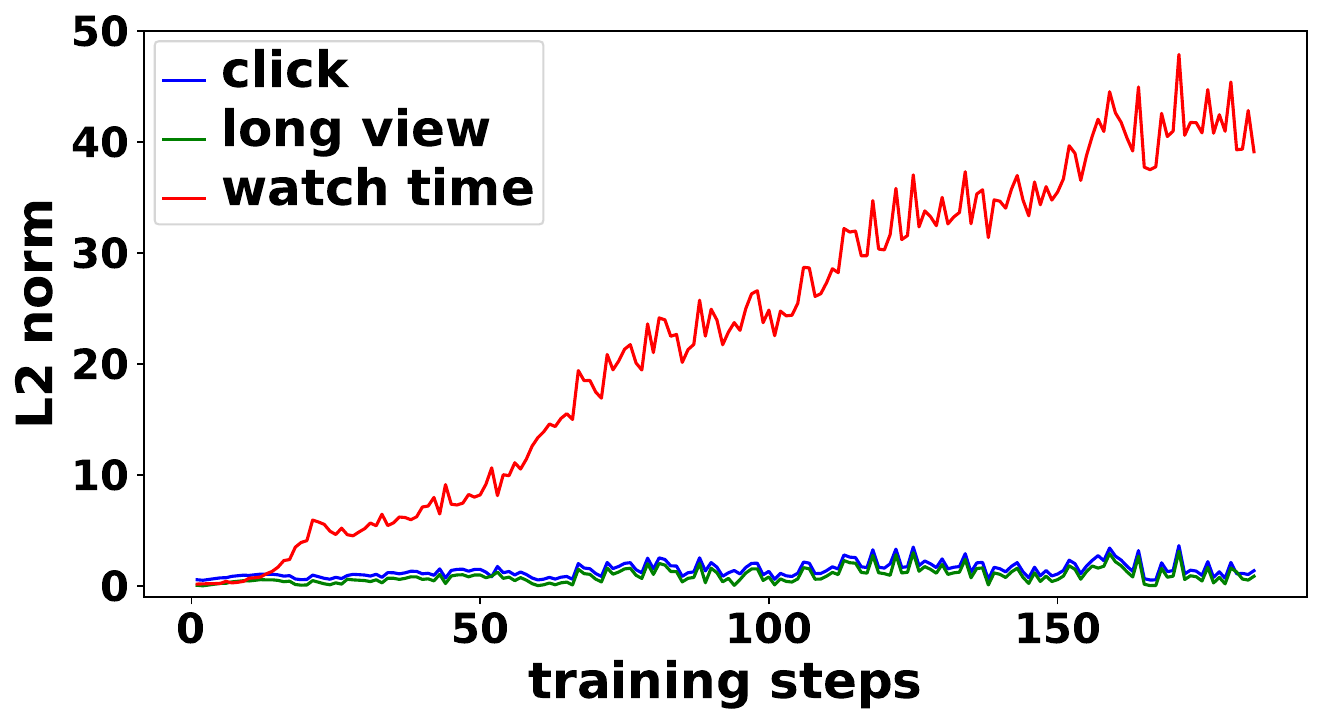}} 
	\subfigure[HTLNet with Gradient Process]{\includegraphics[width=0.75\columnwidth]{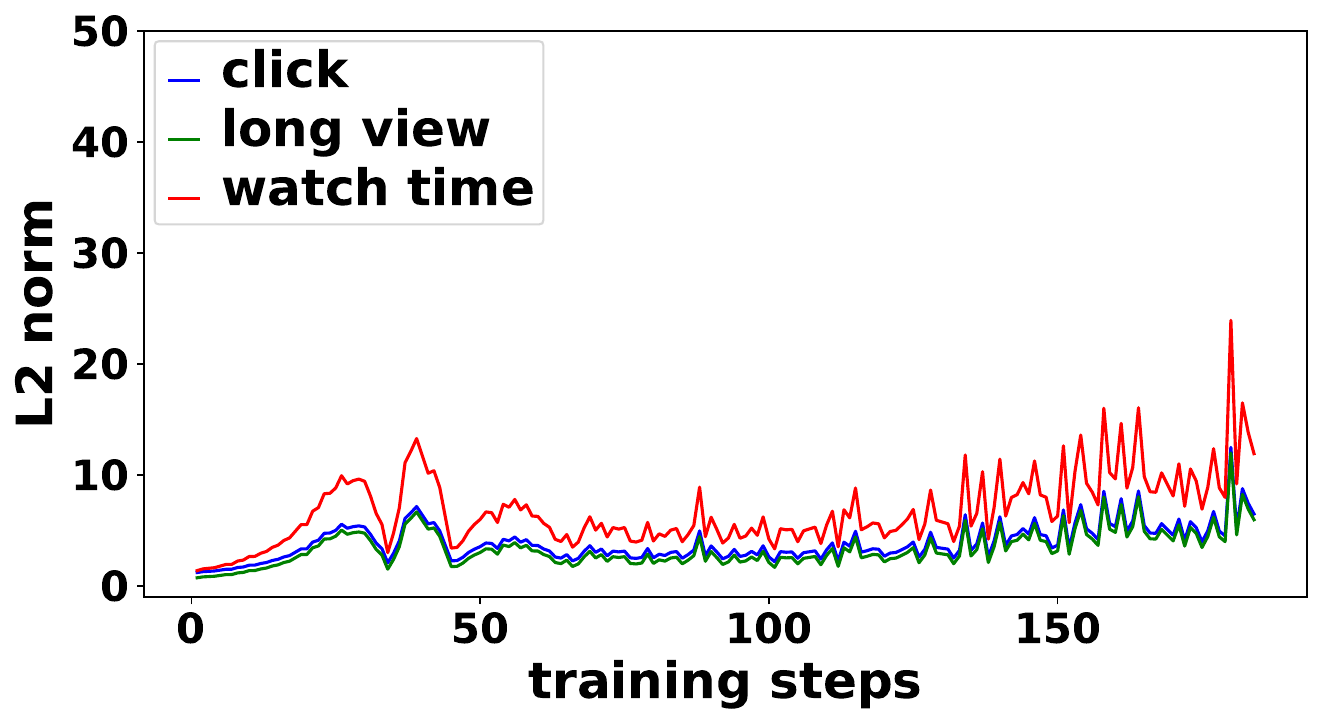}}
    \vspace{-10pt}
	\caption{The illustrative case of the effect of optimization on the KuaiRand dataset. The legend indicates the different tasks. The Y-axis is the average shared embedding gradient magnitude over all mini-batch iterations in one training step.}
    \Description{The illustrative case of the effect of optimization on the KuaiRand dataset.}
    \vspace{-5pt}
\label{fig:comp}
\end{figure}

In addition, we give an illustrative case to analyze the effect of the gradient process on shared embedding. As shown in Figure~\ref{fig:comp}, the magnitudes of gradient w.r.t shared embedding layer of the HTLNet without or with gradient process presents a totally different tendency. The upper one (without the gradient process) shows that the gradient from the \textit{watch time} target is diverging from the \textit{click} and \textit{long view} targets. Meanwhile, the magnitude of \textit{watch time} gradient becomes increasingly more significant than the other two preceding targets. The one with a gradient process can easily control the magnitude of all three tasks, thus making them converge simultaneously. This observation indicates that the convergence of the gradient poses a severe issue in our proposed model and further explains why our model outperforms baselines. Besides, the investigation of two critical parameters $\alpha$ and $\gamma$ related to performance is conducted in Appendix~\ref{ap:param}.

\subsection{Online Experiments (RQ5)}

This section investigates whether HTLNet performs better in the online recommendation scenario. The details of online experiments can be referred to Appendix~\ref{ap:online}.

\begin{figure}[htbp]
    \centering
    \includegraphics[width=0.9\linewidth]{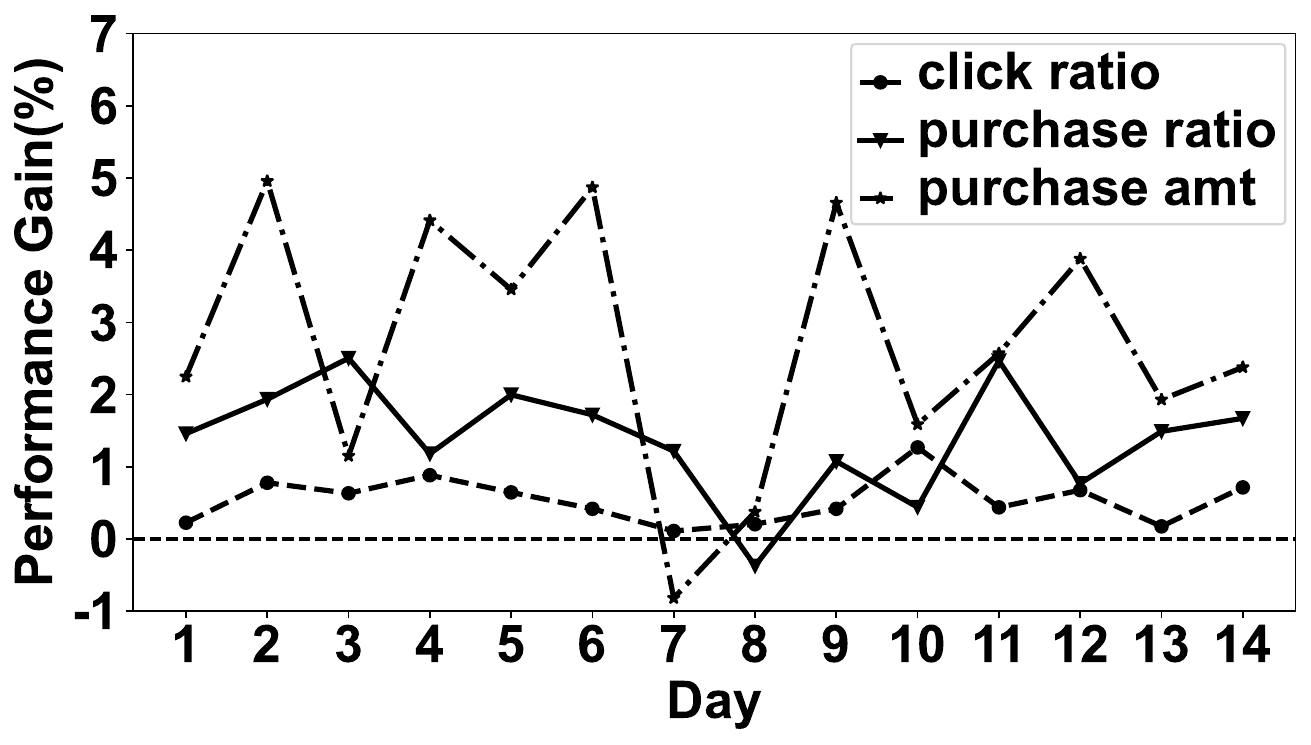}
    \vspace{-10pt}
    \caption{Online relative performance gains of three online metrics in 14 consecutive days.}
    \Description{Online relative performance gains of three online metrics in 14 consecutive days.}
    \label{fig:online}
    \vspace{-5pt}
\end{figure}

Figure~\ref{fig:online} shows the relative promotion of three corresponding objectives in 14 consecutive days. On most days, HTLNet achieves significant performance improvements, especially in terms of purchase amount. The accumulated gains of CTR, CVR, and purchase amount are $+0.54\%$,$+1.4\%$, and $+2.69\%$. These significant improvements in the online scenario prove the effectiveness of HTLNet. However, the purchase amount is considered more fluctuant due to the difficulties in the regression target, which further emphasizes our study.

\section{Conclusion}
In this paper, a novel model, HTLNet, is devoted to learning hybrid targets for the first time. To effectively explore the dependence among the hybrid targets, HTLNet introduces a label embedding unit to map the label into the dense vector containing the information of preceding labels. Then an information fusion unit incorporates preceding information adaptively, which helps the task prediction. Moreover, an optimization strategy is also proposed to solve the gradient issue raised by our network architecture. Compared to several multi-task approaches, the performance gains of HTLNet on both public and product datasets demonstrate its effectiveness in exploring dependence. Besides, we also conduct an online A/B test to further verify our HTLNet performs well in a large-scale fund recommendation scenario.

\paragraph{Limitations}
Despite HTLNet demonstrating superior effectiveness over other baselines, it requires more training times than other baselines, discussed in Appendix~\ref{ap:efficiency}. Additionally, HTLNet posits the hypothesis that the objective of the final core target is to be continuous while those of all the preceding tasks are to be discrete. While this is the most common scenario in the real world, we are interested in generalizing HTLNet into other complex cases.

\begin{acks}
We thank the support of the National Natural Science Foundation of China (No.62302310).
\end{acks}
\bibliographystyle{ACM-Reference-Format}
\bibliography{ref}

\clearpage
\appendix

\section{Motivation for Gumbel-Softmax and its alternative}
\label{ap:gumbel}
In this section, we discuss why the Gumbel-softmax re-parameterization trick~\cite{gumbel} is introduced to differentiate the label information. To understand the gumbel-softmax, the whole process needs to be examined. 

A straightforward strategy is to use the true labels of all predecessor tasks for label transfer. However, since the true labels are available during training but not during testing, this leads to the training-test discrepancy problem.
To this end, proxy labels from the predicted probabilities of the predecessor tasks for label transfer are selected. Simple sampling operations are detrimental to model training because they are non-differentiable. On the one hand, only the label embedding corresponding to one sampled category is updated. This one-sided update quickly leads to high variance in the gradients between different label embeddings and makes the optimization unstable. On the other hand, label transfer based only on information from one category label may be insufficient, especially when the labels of the underlying task are noisy.

To alleviate these problems, we introduce gumbel-softmax to perform a differentiable sampling operation. This means that the model can use the information of all category labels in the early stage to perform inference in label transfer and update the label embeddings of all categories simultaneously. Then, as training progresses, the model gradually and smoothly converges to the final proxy labels and corresponding label embeddings.



\section{Experimental Setup}

\subsection{Offline Implementation Details}
\label{ap:offline}

In our experiments, all models have three hidden layers, the units of which
are $[128, 64, 32]$. Following all previous work, Adam optimizer~\cite{adam}, and Xavier initialization~\cite{xavier} are all adopted. For each model, we tune the following hyper-parameters to get the best performance:
shared embedding dimension, batch size, learning rate, and $l_2$ regularization. For our HTLNet, we set temperature $\tau=10$ and decay with training in $0.5$. We further tune latent label embedding dimension, the decay step for temperature $\tau_d$, $\alpha$, and $\gamma$ in gradient process with the hyper-parameter search library \textit{Optuna}\footnote{https://optuna.org/}. Table~\ref{tab:search_ranges} summarizes the candidate search space for hyper-parameters. In addition, the transferred task representations are extracted from the first layer representation. 

\begin{table}[!htbp]
\caption{Hyper-parameters tuned in the experiments.}
\label{tab:search_ranges}
\vspace{-10pt}
\centering
\resizebox{0.45\textwidth}{!}{
\begin{tabular}{ccc}
\hline
\textbf{Name} & \textbf{Range} & \textbf{Functionality}\\
\hline
$D$ & $\{5,10,15,20\}$ & shared embedding dimension \\
$bs$ &$\{2^{8},2^{9},2^{10},2^{11},2^{12}\}$ & batch size \\
$lr$ &$\{1e^{-5},5e^{-5},1e^{-4},...,5e^{-3},1e^{-2}\}$ & learning rate \\
$\lambda$ &$\{1e^{-5},5e^{-5},1e^{-4},...,5e^{-3},1e^{-2}\}$ & l2 regularization \\
\hline
$L_d$ &$\{5,10,15,20\}$ & label embedding dimension \\
$\tau_d$ &$\{1000, 5000, 10000, 15000, 20000\}$ &  decay step for temperature\\
$\alpha$ &$\{1e^{-6},1e^{-5},1e^{-4},...,1e^{-1},1\}$ & $\alpha$ for gradient conflict \\
$\gamma$ &$\{1e^{-5},5e^{-5},1e^{-4},...,5e^{-1},1\}$ & $\gamma$ for gradient magnitude \\
$clip$ &$\{1,5,1e^{2},5e^{2},...,5e^{3},1e^{4}\}$ & clip for weight \\
\hline \\
\end{tabular}}
\vspace{-10pt}
\end{table}

Our implementation is based on TensorFlow. As some baseline models are implemented on Pytorch, we re-implement them according to the official implementation for the AITM\footnote{https://github.com/xidongbo/AITM}, AdaTT\footnote{https://github.com/facebookresearch/AdaTT} and MetaBalance\footnote{https://github.com/facebookresearch/MetaBalance}. For other baseline models, we re-implement them based on the authors' details, and all experiments are implemented on NVIDIA V100 and Intel(R) Xeon Platinum 8255C CPU@2.50GHz.
\subsection{Evaluation Metric}
\label{ap:metric}
Here we add more detailed description of evaluation metrics. For discrete tasks (mostly preceding tasks), such as \textit{click} and \textit{long view} in KuaiRand dataset, \textit{repurchase 1Y} and \textit{repurchase 1M} in Kaggle-Revenue dataset, \textit{click} and \textit{convert} in Product dataset, the commonly-adopted metrics, AUC and logloss, are adopted. Moreover, for continuous tasks (mainly core tasks), such as \textit{watch time} in KuaiRand dataset, \textit{repurchase 1M amount} in Kaggle-Revenue dataset, and \textit{purchase amount} in Product dataset, normalized root mean squared error (NRMSE) and normalized mean absolute error (NMAE) are adopted. 
Additionally, due to the particularity of the amount, Spearman rank-order correlation coefficient (Spearman) and Gini coefficient (Gini) are adopted for \textit{repurchase 1M amount} in Kaggle-Revenue dataset and \textit{purchase amount} in Product dataset. Both Gini and Spearman are usually used to measure the ranking of predictions in industrial applications~\cite{ziln}.

\subsection{Online Implementation Details}
\label{ap:online}
We first introduce the online scenario and setting and then present the online results.

We conducted an A/B test in the online fund recommendation scenario to measure the benefits of HTLNet compared with the online baseline \textbf{MMoE}. Unlike other recommendation scenarios, the fund recommendation scenario focuses on not only the click/conversion behaviors but also the users' purchase amount on the platform.
We allocate 10\% serving traffic for 14 days. 
Both models are trained in a single cluster, where each node contains a 96-core Intel(R) Platinum 8255C CPU, 256GB RAM, and 8 NVIDIA TESLA A100 GPU cards. Three online metrics, i.e., click ratio (CTR), purchase ratio (CVR), and purchase amount (core target), are measured in the online performance.
\section{Additional Experiments}

\subsection{HTLNet with Lognormal Loss}
\label{ap:lognormal}
To handle this special target, we compare ZILN and HTLNet with ZILN architecture on these two datasets as shown in Table~\ref{tab:ziln_result}. Compared with MSE loss in Table~\ref{tab:main_result_both}, the lognormal loss improves the performance of all the tasks, which indicates that the lognormal loss is a better inductive bias for revenue prediction. Notice that our HTLNet still performs better with Ziln, indicating the superiority and generalization of our HTLNet on hybrid target learning.
\begin{table}[!htbp]
\caption{Results of HTLNet with ZILN Loss on Kaggle-Revenue and product dataset}
\vspace{-10pt}
\centering
\resizebox{0.45\textwidth}{!}{
\begin{tabular}{c|c|c|c|c}
\hline
     Dataset & Task & Metrics & ZILN & HTLNet-ZILN \\ 
\hline
    \multirow{7}*{Kaggle-} 
    & \multirow{2}*{repurchase 1Y}
         & AUC$\uparrow$ & - & $\textbf{0.7022}^{*}$ \\
    \multirow{7}*{Revenue} 
    &     & Logloss$\downarrow$ & - & $\textbf{0.3780}^{*}$ \\
\cline{2-5}
    & \multirow{2}*{repurchase 1M} 
         & AUC$\uparrow$ & 0.6320 & $\textbf{0.6348}^{*}$ \\ 
    &     & Logloss$\downarrow$ & 0.6507 & $\textbf{0.6499}$ \\
\cline{2-5}
    & \multirow{3}*{repurchase 1M} 
         & NRMSE$\downarrow$ & 0.8832 & $\textbf{0.8540}^{*}$ \\ 
    & \multirow{3}*{amount} 
         & NMAE$\downarrow$ & 1.1463 & $\textbf{1.1447}$ \\
    &     & Spearman$\uparrow$ & 0.2525 & $\textbf{0.2568}^{*}$ \\ 
    &     & Gini$\uparrow$ & 0.5304 & $\textbf{0.5341}^{*}$ \\ 
    
\hline
    \multirow{8}{*}{Product} 
    & \multirow{2}{*}{click}
         & AUC$\uparrow$ & - & $\textbf{0.8091}^{*}$ \\
    &     & Logloss$\downarrow$ & - & $\textbf{0.1466}^{*}$ \\
\cline{2-5}
    & \multirow{2}{*}{convert} 
         & AUC$\uparrow$ & 0.9041 & $\textbf{0.9082}^{*}$ \\ 
    &     & Logloss$\downarrow$ & 0.1024 & $\textbf{0.1021}$ \\
\cline{2-5}
    & \multirow{3}{*}{purchase} 
         & NRMSE$\downarrow$ & 0.8884 & $\textbf{0.8693}$ \\ 
    & \multirow{3}{*}{amount}
         & NMAE$\downarrow$ & 0.8794 & $\textbf{0.8570}$ \\
    &     & Spearman$\uparrow$ & 0.2444 & $\textbf{0.2547}^{*}$ \\ 
    &     & Gini$\uparrow$ & 0.8691 & $\textbf{0.8749}^{*}$ \\ 
\hline
\end{tabular}}
\label{tab:ziln_result}
\end{table}


\subsection{Efficiency Analysis}
\label{ap:efficiency}

\begin{figure}[!htbp]
    \centering
    \includegraphics[width=0.95\linewidth]{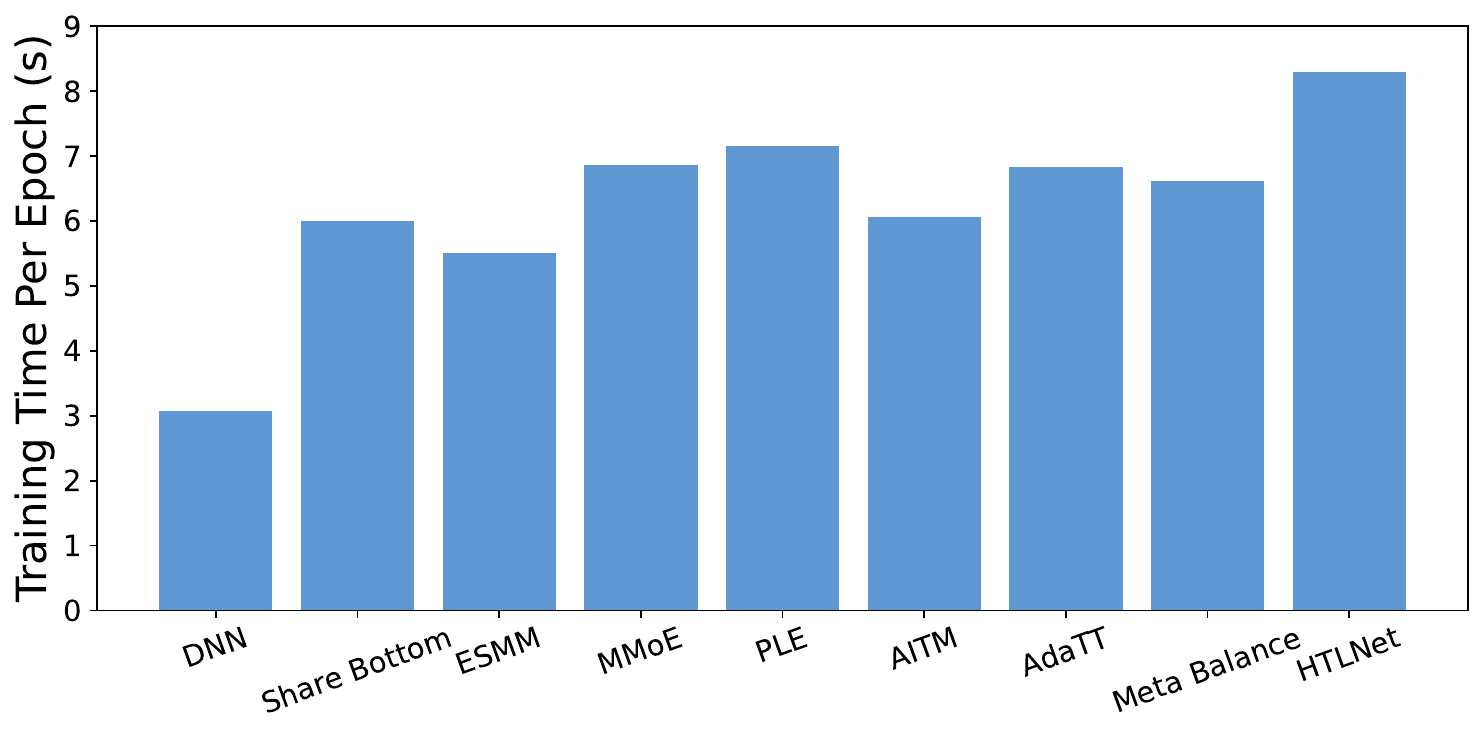}
    \vspace{-10pt}
    \caption{Efficiency Analysis on KuaiRand dataset.}
    \Description{Efficiency Analysis on KuaiRand dataset.}
    \vspace{-10pt}
    \label{fig:efficiency}
\end{figure}

We conduct a training efficiency analysis of different models in Figure~\ref{fig:efficiency}. Here, we train each model over the KuaiRand dataset for 20 epochs, rerun this process ten times, and report the mean training time of each epoch. We observe that HTLNet requires relatively more training time than other baselines.

\subsection{Parameters Sensitivity}
\label{ap:param}

\begin{figure}[!htbp]
	\centering
	\subfigure[NRMSE]{\includegraphics[width=0.8\columnwidth]{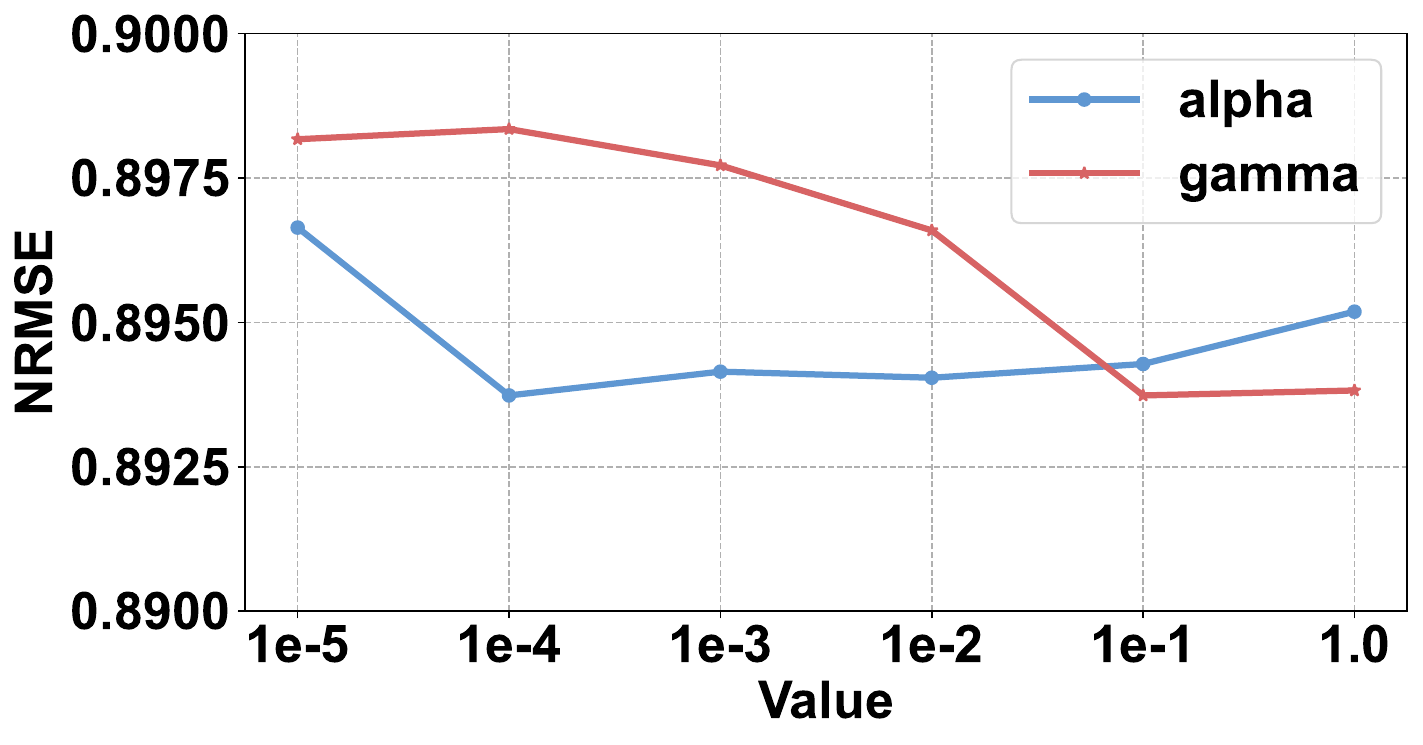}}
	\subfigure[NMAE]{\includegraphics[width=0.8\columnwidth]{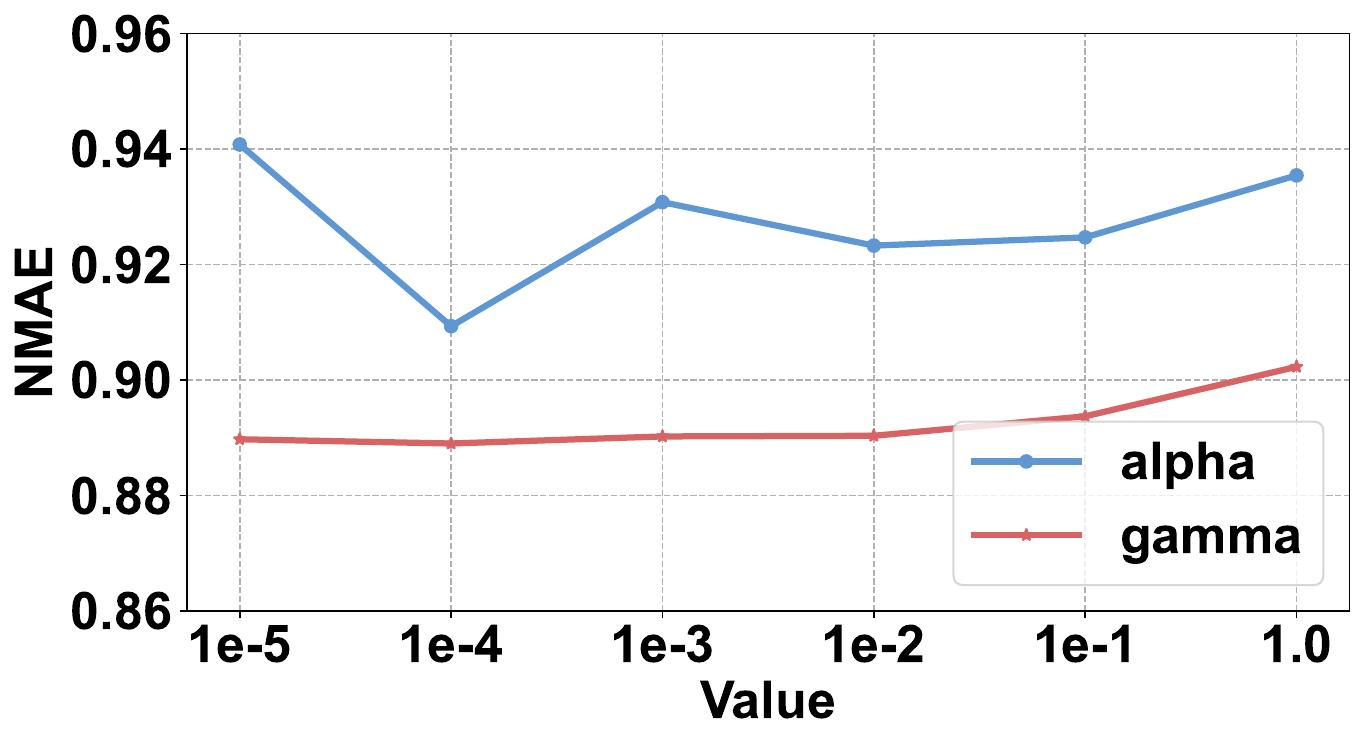}}
    \vspace{-10pt}
	\caption{Parameter study of optimization hyperparameters $\alpha$ and $\gamma$ on NRMSE and NMAE of KuaiRand Dataset.}
    \Description{Parameter study of optimization hyperparameters $\alpha$ and $\gamma$ on NRMSE and NMAE of KuaiRand Dataset.}
\label{fig:param}
\end{figure}

We conduct the parameter sensitivity on the two critical hyperparameters in the gradient process of HTLNet, i.e. $\alpha$ and $\gamma$, which affects the optimization strategy. The results on the KuaiRand dataset are shown in Figure~\ref{fig:param}. 

In Figure~\ref{fig:param}, we varies the both $\alpha$ and $\gamma$ in the range $[1e^{-5}, 1]$ with the step of $10$. Overall, increasing both $\alpha$ and $\gamma$ will improve the performance at the beginning, but too large a value will degrade the performance. However, the effect of $\gamma$ is less sensitive compared with $\alpha$. We speculate that the gradients of the core target magnitude differ from preceding targets in a certain value, far way from which will be detrimental to the core target performance. While $\alpha$ controlling the direction will be smoother in the tuning process.

\end{document}